\documentclass[unsortedaddress,
twocolumn,showpacs,preprintnumbers,amsmath,amssymb]{revtex4}
\usepackage{graphicx}
\usepackage{dcolumn}
\usepackage{bm}
\usepackage{latexsym}
\begin{document}
\title{Cluster formation and anomalous fundamental diagram
in an ant trail model
}
\author{Katsuhiro Nishinari}
\email{kn@thp.uni-koeln.de}
\thanks{Permanent address: Department of Applied Mathematics and Informatics, 
Ryukoku University, Shiga, Japan (email: {\tt knishi@rins.ryukoku.ac.jp})}
\affiliation{%
Institut  f\"ur Theoretische  Physik, Universit\"at zu
K\"oln D-50937 K\"oln, Germany
}%
\author{Debashish Chowdhury}
\email{debch@iitk.ac.in} 
\affiliation{%
Department of Physics, Indian Institute of Technology,
Kanpur 208016, India.
}%
\author{Andreas Schadschneider}%
 \email{as@thp.uni-koeln.de}
\affiliation{%
Institut  f\"ur Theoretische  Physik, Universit\"at 
zu K\"oln D-50937 K\"oln, Germany
}%

\date{\today}%
\begin{abstract} 
A recently proposed stochastic cellular automaton model 
({\it J. Phys. A 35, L573 (2002)}), motivated by the motions of ants 
in a trail, is investigated in detail in this paper. The flux of ants 
in this model is sensitive to the probability of evaporation of 
pheromone, and the average speed of the ants varies non-monotonically 
with their density. This remarkable property is analyzed here 
using phenomenological and microscopic approximations 
thereby elucidating the nature of the spatio-temporal 
organization of the ants.   We find that the observations can be
understood by the formation of loose clusters, i.e.\ space
regions of enhanced, but not maximal, density.
\end{abstract}
\maketitle
\section{\label{sec1}Introduction}
Particle-hopping models have been used widely in the recent years 
to study the spatio-temporal organization in systems of interacting 
particles driven far from equilibrium 
\cite{sz,schutz,privman,derrida,md,droz}. 
Often such models are formulated in terms of cellular automata (CA) 
\cite{wolfram}. Examples of such systems include vehicular traffic 
\cite{css,helbing,nagatani,tgf}
where the vehicles are represented by particles while their mutual 
influence is captured by the inter-particle interactions. Usually, 
these inter-particle interactions tend to hinder their motions so
that the {\it average speed} decreases {\it monotonically} with the
increasing density of the particles. In the usual form of the
fundamental diagram, i.e.\ the flux-density relation, this 
non-monotonicity corresponds to the existence of an inflection point.
In a recent letter \cite{cgns} we have 
reported a counter-example, motivated by the flux of ants in a trail
\cite{burd}, where, the average speed of the particles varies
{\it non-monotonically} with their density because of the coupling
of their dynamics with another dynamical variable. In 
\cite{cgns} we presented numerical evidences in support of 
this unusual feature 
of the model and indicated the physical origin of this behaviour in 
terms of a heuristic mean-field argument. In this paper we present 
the corresponding detailed {\it analytical} calculations, together 
with further numerical results, that provide deep insight into the model. 

The paper is organized as follows: The ant-trail model \cite{cgns} 
is defined in Section \ref{sec2} and compared with some closely 
related models in Section \ref{sec3}. 
Sec.~\ref{secClust} presents results obtained from a microscopic
cluster approximation. Although this approach does not reproduce
the observed sharp crossover it will help us to get a better understanding
of the microscopic structure of the stationary state.
A heuristic homogeneous mean-field theory, which was  
sketched briefly in ref.~\cite{cgns}, is presented in detail 
in Section \ref{sec4}. This theory produces better results than
the cluster approximation. However, it accounts only for the 
{\it qualitative} features of the fundamental diagram obtained by computer 
simulations. Therefore in Section \ref{sec5} we present a new approach
which leads to the main new results. 
In this section we have computed some new quantities that provide 
information as to the state of occupation of the site immediately in 
front of an ant. These quantities not only help us in identifying {\it 
three} regimes of density, with corresponding characteristic features, 
but also provide insights that we exploit in developing a new scheme 
for analytical calculations. The results of this new scheme, that we 
call ``loose-cluster approximation'' (for reasons which will be clear in 
Section \ref{sec5}), are in reasonably good {\it quantitative} agreement with 
the data obtained from computer simulations. The effects of replacing the 
parallel updating by random sequential updating is explored in Section 
\ref{sec6}. 
The results are summarized and conclusions are drawn in Section \ref{sec7}. 
In two appendices some details of the calculations for
the cluster-theoretic approaches are given.

\section{\label{sec2}The ant-trail model}
The ants  communicate with each other by dropping a chemical
(generically called {\it pheromone}) on the substrate as they crawl
forward \cite{wilson,camazine}. Although we cannot smell it, the trail
pheromone sticks to the substrate long enough for the other following
sniffing ants to pick up its smell and follow the trail. 
Ant trails may serve different purposes (trunk trails, migratory routes)
and may also be used in a different way by different species.
Therefore one-way trails are observed as well as trails with counterflow
of ants.

In \cite{cgns} we developed a particle-hopping model, formulated in 
terms of stochastic CA \cite{wolfram}, which may be interpreted as a 
model of unidirectional flow in an ant-trail. 
As in ref.~\cite{cgns}, rather than addressing the question of the 
emergence of the ant-trail, we focus attention here on the traffic of 
ants on a trail which has already been formed. Furthermore we have
assumed unidirectional motion. The effects of counterflow, which are 
important for some species, will be investigated in the future.

\begin{figure}[tb]
\begin{center}
\includegraphics[width=0.45\textwidth]{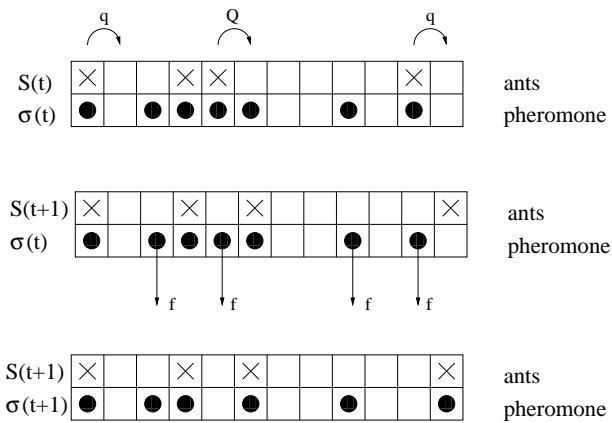} 
\end{center}
\caption{
Schematic representation of typical configurations; it
also illustrates the update procedure. Top: Configuration at time $t$,
i.e.\ {\it before} stage $I$ of the update. The non-vanishing hopping
probabilities of the ants are also shown explicitly. Middle:
Configuration {\it after} one possible realisation of {\it stage $I$}.
Two ants have moved compared to the top part of the figure. Also
indicated are the pheromones that may evaporate in stage $II$ of the
update scheme.  Bottom: Configuration {\it after} one possible realization
of {\it stage $II$}. Two pheromones have evaporated and one pheromone has
been created due to the motion of an ant.
}
\label{fig-0}
\end{figure}

Each site of our one-dimensional ant-trail model represents a cell
that can accomodate at most one ant at a time (see Fig.~\ref{fig-0}).
The lattice sites are labelled by the index $i$ ($i = 1,2,...,L$);
$L$ being the length of the lattice. We associate two binary variables
$S_i$ and $\sigma_i$ with each site $i$ where $S_i$ takes the value
$0$ or $1$ depending on whether the cell is empty or occupied by an ant.
Similarly, $\sigma_i =  1$ if the cell $i$ contains pheromone; otherwise,
$\sigma_i =  0$. Thus, we have two subsets of dynamical variables in
this model, namely,
$\{S(t)\} \equiv (S_1(t),S_2(t),...,S_i(t),...,S_L(t))$ 
and
$\{\sigma(t)\} \equiv (\sigma_1(t),\sigma_2(t),...,\sigma_i(t),...,
\sigma_L(t))$.
The instantaneous state (i.e., the configuration) of the system at
any time is specified completely by the set $(\{S\},\{\sigma\})$.

Since a unidirectional motion is assumed, ants do not move backward.
Their forward-hopping probability is higher if it smells pheromone 
ahead of it.
The state of the system is updated at each time step in {\it two
stages}.  In stage I ants are allowed to move. Here the subset
$\{S(t+1)\}$ at the time step $t+1$ is obtained using the full information
$(\{S(t)\},\{\sigma(t)\})$ at time $t$. Stage II corresponds to the
evaporation of pheromone. Here only the subset $\{\sigma(t)\}$ is
updated so that at the end of stage II the new configuration
$(\{S(t+1)\},\{\sigma(t+1)\})$ at time $t+1$ is obtained.
In each stage the dynamical rules are applied {\it in parallel} to
all ants and pheromones, respectively.\\

\noindent {\it Stage I: Motion of ants}\\[0.2cm]
\noindent An ant in cell $i$ that has an empty cell in front of it, i.e., 
$S_i(t)=1$ and $S_{i+1}(t)=0$, hops forward with
\begin{equation}
{\rm probability} = \left\{\begin{array}{lll}
            Q &\quad{\rm if\ }~\sigma_{i+1}(t) = 1,\\
            q &\quad{\rm if\ }~\sigma_{i+1}(t) = 0,
\end{array} \right.
\end{equation}
where, to be consistent with real ant-trails, we assume $ q < Q$.\\

\noindent {\it Stage II: Evaporation of pheromones}\\[0.2cm]
\noindent At each cell $i$ occupied by an ant after stage I
a pheromone will be created, i.e., 
\begin{equation}
\sigma_i(t+1) = 1\quad {\rm if\ }\quad S_i(t+1) = 1.
\end{equation}
On the other hand, any `free' pheromone at a site $i$ not occupied
by an ant will evaporate with the probability $f$ 
per unit time, i.e., if $S_i(t+1) = 0$, $\sigma_i(t) = 1$, then
\begin{equation}
\sigma_i(t+1) = \left\{\begin{array}{lll}
0 &\quad {\rm with\ probability\ } f,\\
1 &\quad {\rm with\ probability\ } 1-f.
\end{array} \right.
\end{equation}

Note that the dynamics conserves the number $N$ of ants, but not
the number of pheromones.

The rules can be written in a compact form as 
the coupled equations
\begin{eqnarray}
 S_j(t+1)&=&S_j(t)+\min(\eta_{j-1}(t),S_{j-1}(t),1-S_j(t))\nonumber\\
&&\hspace{0.5cm}-\min(\eta_{j}(t),S_{j}(t),1-S_{j+1}(t)),\label{eqa}\\
\sigma_j(t+1)&=&\max(S_j(t+1),\min(\sigma_j(t),\xi_j(t))),\label{eqf}
\end{eqnarray}
where $\xi$ and $\eta$ are stochastic variables defined by 
$\xi_j(t)=0$ with the probability $f$ and $\xi_j(t)=1$ with $1-f$, 
and $\eta_j(t)=1$ with the probability $p=q+(Q-q)\sigma_{j+1}(t)$ and 
$\eta_j(t)=0$ with $1-p$. This representation is useful for the
development of approximation schemes.

\section{\label{sec3}Comparison with other models}
In this section we compare the ant-trail model first with the 
Nagel-Schreckenberg (NS) model \cite{ns} to show that in various 
limits it reduces to the NS model with different hopping probabilities.
This comparison also helps in formulating the task of our analytical 
calculation from alternative perspectives. We also compare the 
ant-trail model with some other models all of which share a common 
feature: the dynamics of the ``particles'' are coupled to another 
dynamical variable.
 
\subsection{The Nagel-Schreckenberg model}
The NS model \cite{ns} is the minimal particle-hopping model for 
vehicular traffic on freeways. In the general version of the NS 
model the particles, each of which represents a vehicle, can have 
a maximum speed of $V_{{\rm max}}$. However, by the term 'NS model' in 
this paper we shall always mean the NS model with $V_{{\rm max}} = 1$, 
so that each particle can move forward, by one lattice spacing, 
with probability $q_{NS}$ if the lattice site immediately in front 
is empty. 

The most important quantity of interest in the context of flow
properties of the traffic models is the {\it fundamental diagram},
i.e., the flux-versus-density relation, where flux is the product of
the density and the average speed. For a hopping probability $q_{NS}$
at a given density $\rho=N/L$ the exact flux $F(\rho)$ in the NS model 
is given by \cite{css,ssni}
\begin{equation}
F_{NS}(\rho)= \frac{1}{2}\left[1-\sqrt{1 ~- ~4 ~q_{NS} ~\rho(1-\rho)}\right]
\label{eq-nsflux}
\end{equation}
which reduces to $F_{NS}(\rho) = \min(\rho,1-\rho)$ 
in the deterministic limit $q_{NS} = 1$. 

\begin{figure}[tb]
\begin{center}
\includegraphics[width=0.4\textwidth]{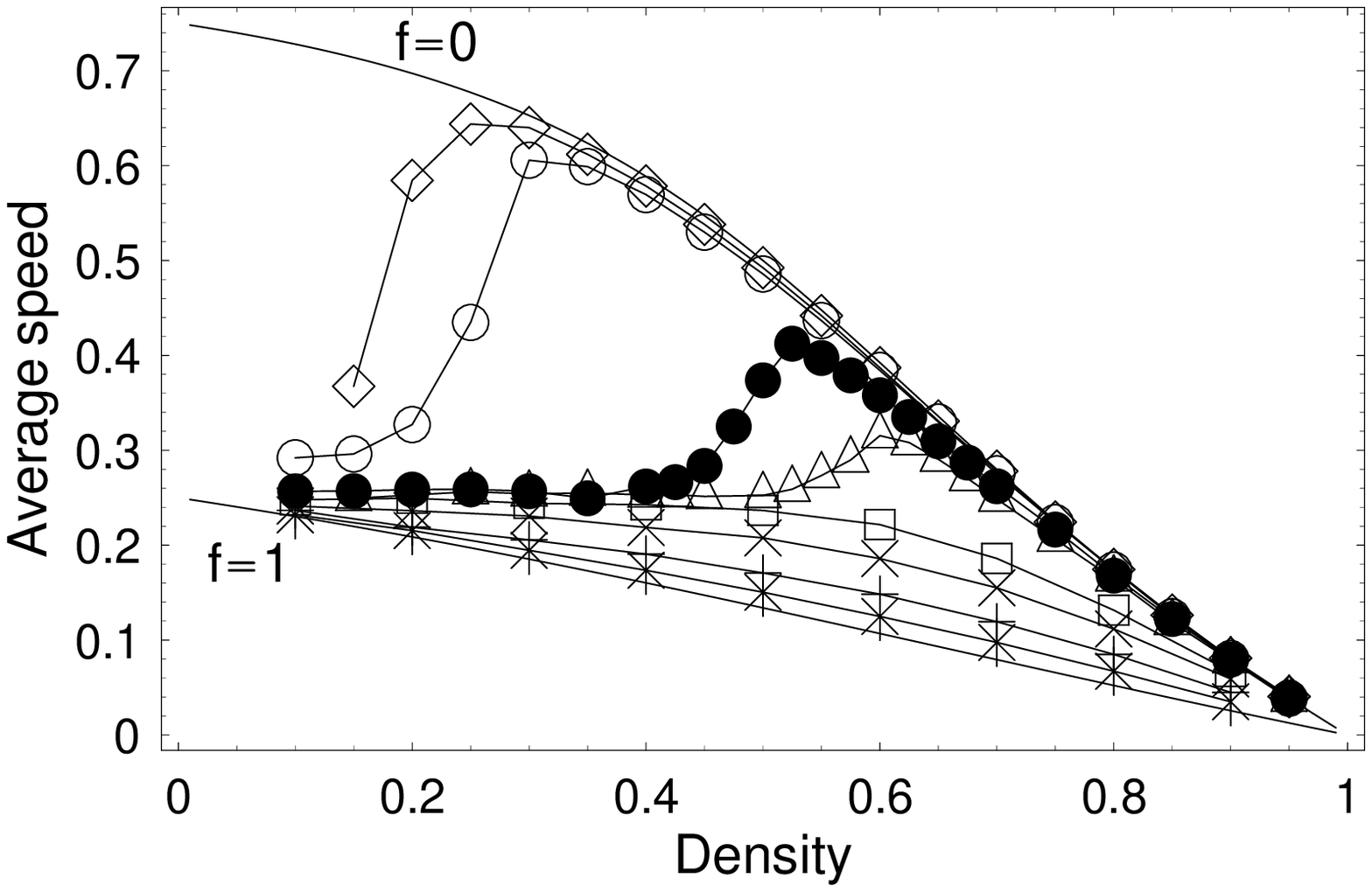}\\
\hspace{1cm}(a)\\
\vspace{0.5cm}
\includegraphics[width=0.4\textwidth]{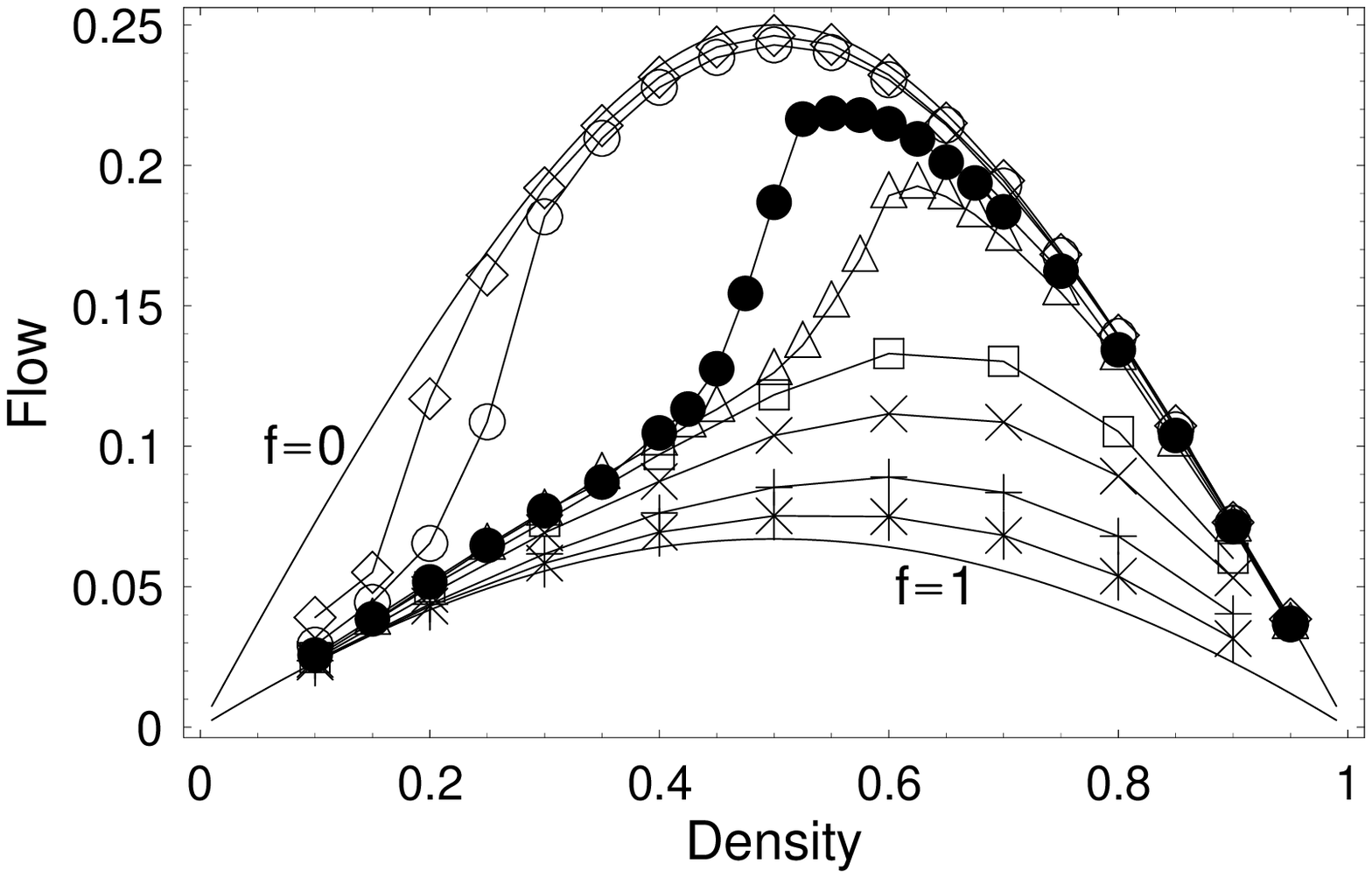}\\
\hspace{1cm}(b)\\
\vspace{0.5cm}
\includegraphics[width=0.4\textwidth]{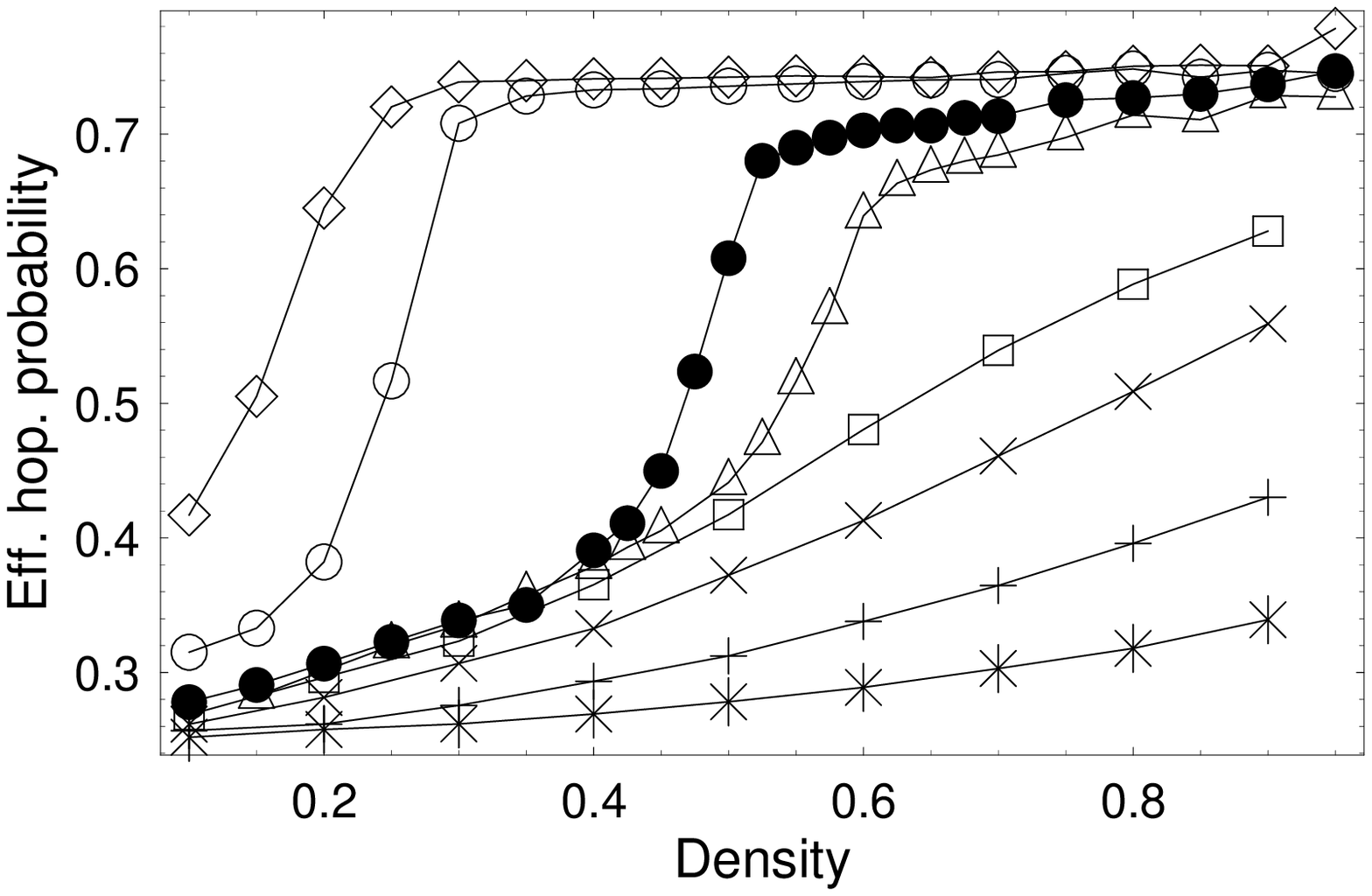}\\
\hspace{1cm}(c)
\end{center}
\caption{The average speed (a), flux (b) and effective hopping probability
(c) of the ants, extracted from computer simulation data, are plotted against
their densities for the parameters $Q = 0.75, q = 0.25$.
The discrete data points corresponding to $f=0.0005 ({\Diamond})$,
$0.001 (\circ$), $0.005 (\bullet)$, $0.01 ({\bigtriangleup})$,
$0.05 ({\Box})$, $0.10 (\times)$, $0.25 (+)$, $0.50 (\ast)$ have been
obtained from computer simulations; the dotted lines connecting these
data points merely serve as the guide to the eye. In (a) and (b), the
 cases $f=0$ and $f=1$ are also displayed, which correspond to the NS
 model
with $q_{{\rm eff}}=Q$ and $q$, respectively.}
\label{fig-effh}
\end{figure}

Note that the expression (\ref{eq-nsflux}) remains invariant under the 
interchange of $\rho$ and $1-\rho$; this ``particle-hole'' symmetry of the 
NS model leads to a fundamental diagram that is symmetrical about 
$\rho = \frac{1}{2}$. In contrast, the fundamental diagrams of our 
ant-trail model (see Fig.~\ref{fig-effh}(b)) do not possess this 
symmetry except in the special cases of $f = 0$ and $f = 1$. As explained in 
ref.~\cite{cgns}, in the two special cases $f = 0$ and $f = 1$ the 
ant-trail model becomes identical to the NS model with $q_{NS} = Q$ 
and $q_{NS} = q$, respectively, and, hence, recovers the particle-hole 
symmetry in these two special limits. 

The flux $F$ and the average speed $V$ of vehicles are related
by the hydrodynamic relation $F=\rho V$. The density-dependence of the 
average speed in our ant-trail model is shown in Fig.~\ref{fig-effh}(a). 
Over a range of small values of $f$, it exhibits an anomalous behaviour
in the sense that, unlike common vehicular traffic, $V$ is not a 
monotonically decreasing function of the density $\rho$.
Instead a relatively sharp crossover can be observed where the
speed {\em increases} with the density. In the usual form of the
fundamental diagram (flux vs.\ density) this transition leads to
the existence of an inflection point (Fig.~\ref{fig-effh}(b)).

Assuming that the flux in ant-trail model is given by the equation 
(\ref{eq-nsflux}) with a an effective hopping probability 
$q_{{\rm eff}}(\rho)$, which depends on the ant density $\rho$, we can 
extract $q_{{\rm eff}}(\rho)$ by fitting the observed flux 
with $F_{NS}(\rho)$, i.e., from
\begin{equation}
q_{\rm eff} = \frac{F(1-F)}{\rho(1-\rho)}.
\end{equation}
The effective hopping probability, $q_{{\rm eff}}$, is plotted as a function
of $\rho$ for several different values of the parameter $f$ in 
Fig.~\ref{fig-effh}(c). 
In the limit $\rho \rightarrow 0$, the pheromone dropped
by an ant gets enough time to completely evaporate before the
following ant comes close enough to smell it; therefore the ants'
hopping probability is almost always $q$. On the other hand, in
the opposite limit $\rho \rightarrow 1$, the ants are too close to
miss the smell of the pheromone dropped by the leading ant unless
the pheromone evaporation probability is very high; consequently,
in the limit the ants hop most often with the probability $Q$.

A proper theory of the ant-trail model should reproduce the 
non-monotonic variation of the average speed with density (shown in
Fig.~\ref{fig-effh}(a)) and, hence, the unusual shape of the fundamental 
diagram (shown in Fig.~\ref{fig-effh}(b)).
In this paper we develope theories which, indeed, reproduce these features.

\subsection{Models with coupled dynamical variables}
Models with coupled dynamical variables have been considered earlier,
for example, in the context of reaction-controlled diffusion \cite{trimper}. 
However, in this subsection we compare the ant-trail model with 
some more closely related models where the movement of the 
``particles'' is totally asymmetric. 

In the ant-trail model developed 
in ref.~\cite{helbing2} the particles, which represent the ants, move 
in a ``ground-potential landscape'' created by the pheromones. A 
similar approach has also been used for studying 
the human-trails of pedestrians \cite{helbing2}.

In the CA model introduced in ref.~\cite{schad} for pedestrian
dynamics, the floor-fields, albeit {\it virtual}, are
analogs of the pheromone fields $\{\sigma\}$ in the
ant-trail model. 
However, these floor-fields are ``bosonic'' in the sense that the 
variable $\sigma$, which is by definition non-negative, has an 
otherwise unrestricted range.
In contrast, in our ant-trail model the pheromone field is
``fermionic'' as the variable $\sigma$, representing pheromones, can 
take only two values, namely, $0$ (absence) and $1$ (presence). 

The ant-trail model we propose here is closely related to the bus route
model \cite{loan} with parallel updating \cite{cd}. In fact, as we'll
argue now, the ant-trail model and the bus route model are the two
opposite limits of the same generalized version of the NS model of
vehicular traffic. The ants are the analogs of the buses while the
cells accomodating ants in the ant-trail model are analogs of the bus
stops in the bus-route model. Both the models involve two dynamical
variables; the variables $S$ and $\sigma$ in the ant-trail model are
the analogs of the variables representing the presence (or absence)
of bus and passengers, respectively, in the bus route model.
Just as the number of buses is conserved in the bus-route model, the
number of ants is also conserved in our ant-trail model. Similarly, the
dynamical variable representing  the presence (or absence) of pheromone
is not conserved in the ant-trail model just as the number of passangers
is not conserved by the dynamics of the bus-route model. However, there
is a crucial difference between these two models; in the bus-route
model $Q < q$ (as the buses must {\it slow down} to pick up the waiting
passengers) whereas in our ant-trail model $Q > q$ (because an ant is
more likely to move forward if it smells pheromone ahead of it).
In addition, the pheromone are {\it dropped} by ants (whereas passengers 
arrive at the bus stops independent of the buses) while passengers 
are {\it picked up} by buses (whereas pheromones evaporate independently).

\section{\label{secClust} Cluster approximation} 

The simulation results indicate that correlations between different
ants as well as between ants and pheromone play an important role. 
We therefore develop a microscopic ``(2+1)``-cluster approximation 
\cite{css,ssni,kolmo} which 
allows the inclusion of correlations between the occupation variables 
$S_{j-1}(t)$ and $S_j(t)$ of two successive sites $j-1$ and $j$ 
(corresponds to ``2'') and that between the variables $S_j(t)$ 
and $\sigma_j(t)$ at the same site $j$ (corresponds to ``1'')
in an exact way. 

The central quantities of the (2+1)-cluster approximation are the
eight variables
\begin{eqnarray}
P(S_{j-1}(t)S_j(t)), \qquad
P\left(\begin{array}{c}
S_j(t) \\ \sigma_j(t)
\end{array}\right),
\label{PSsigma}
\end{eqnarray}
corresponding to all possibilities 
($S_{j-1}(t),S_j(t),\sigma_j(t)\in \{0,1\}$) of finding the 
corresponding configurations of $S$ and $\sigma$ at a time.
In Appendix~\ref{AppendA} we will show how the master equation
for these quantities can be derived from microscopic considerations
and how the resulting equations can be solved consistently.

\begin{figure}[tb]
\begin{center}
\includegraphics[width=0.45\textwidth]{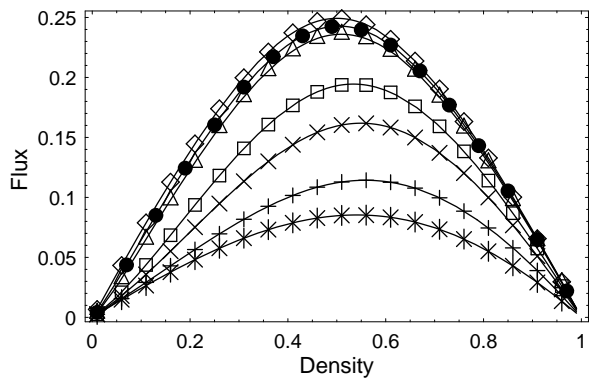}\\
\end{center}
\caption{Fundamental diagrams in the $(2+1)$-cluster 
approximation. The hopping probabilities are $Q=0.75,q=0.25$. 
The same symbols in Fig.~\ref{fig-effh} and in this figure correspond to 
the same values of $f$.
}
\label{fig-A1}
\end{figure}

The flux is given by
\begin{equation}
F = q_{{\rm eff}} P(10).
\label{defF}
\end{equation}
In Appendix~\ref{AppendA} it is shown that within the (2+1)-cluster 
approximation considered here, $F$ can be obtained from the solution 
of the cubic equation
\begin{equation}
 F^2-F+\rho(1-\rho)\left\{
q+\frac{(Q-q)(1-f)F}{(1-\rho)f+(1-f)F}
\right\}=0.
\label{flowfin}
\end{equation}
The result is shown in Fig.~\ref{fig-A1}.
For all values of $f$ in the range $ 0 < f < 1$, the peak of the flux 
appears at $\rho > 1/2$, in qualitative agreement with the general trend 
observed in Fig.~\ref{fig-effh}. But, 
this (2+1)-cluster approximation cannot reproduce the sharp rise in 
the fluxes observed in Fig.~\ref{fig-effh}. 
Note that in each of the three cases $Q=q$, $f=0$ and $f=1$
 the solution of
(\ref{flowfin}) is identical to (\ref{eq-nsflux}) with either 
$q_{NS}=q$ or $q_{NS}=Q$.

Next let us define $P(m)$ as the probability of finding $m$-size cluster 
of {\em ants} in a stationary state. Here the 1-size cluster is defined by
$\cdots 010 \cdots$, and a $m$-size cluster consists of
a string of $m$ of ``1'' between ``0''s. The distribution of cluster
sizes is then given by (see Appendix~\ref{AppendA})
\begin{equation}
 P(m)=\frac{P(10)}{\rho}\frac{\left(1-\frac{P(10)}{\rho}\right)^{m-1}}{
1-\left(1-\frac{P(10)}{\rho}\right)^{\rho L}}.
\label{pm2}
\end{equation}
\begin{figure}[t]
\begin{center}
\includegraphics[width=0.23\textwidth]{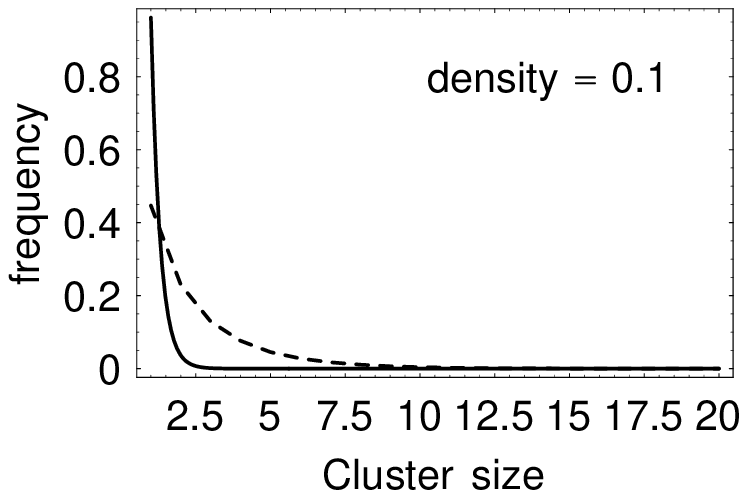}
\includegraphics[width=0.23\textwidth]{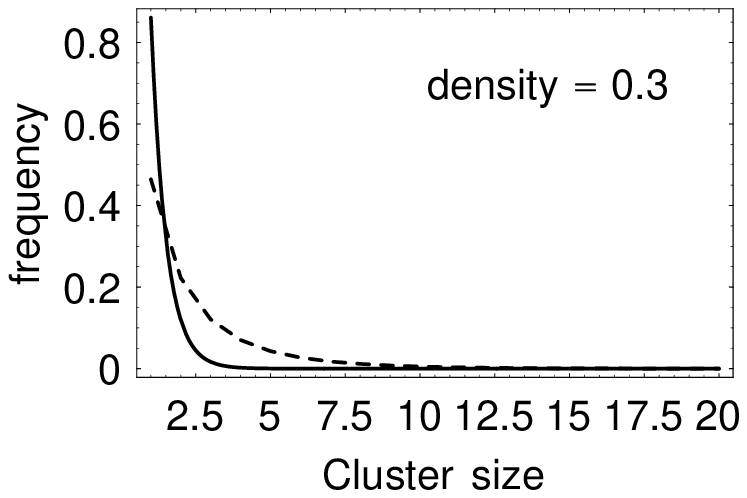}\\
\includegraphics[width=0.23\textwidth]{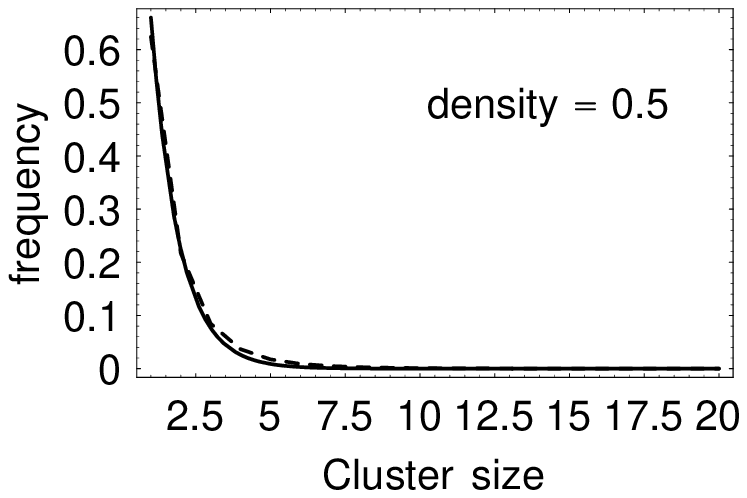}
\includegraphics[width=0.23\textwidth]{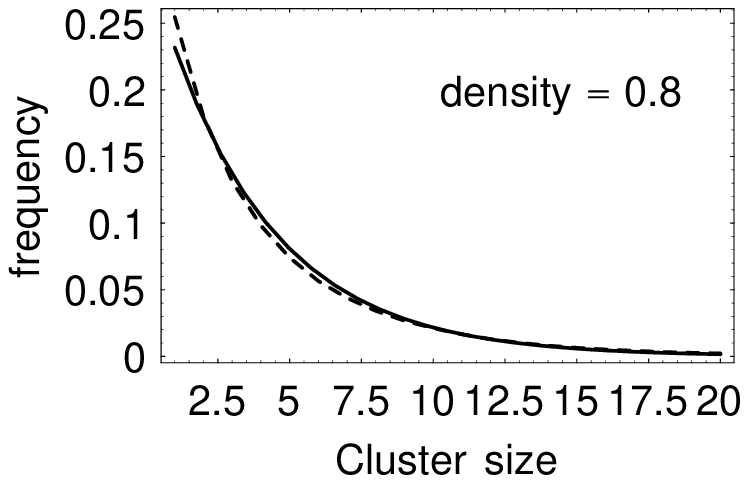}\\
\end{center}
\caption{Cluster-size distribution for $\rho=0.1,0.3,0.5,0.8$.
Theoretical curve (solid line) given by (\ref{pm2}) 
underestimates the simulation (broken curve) at densities $\rho < 0.5$. 
Other relevant parameters are $Q=0.75, q=0.25, f=0.005$.}
\label{figA2}
\end{figure}

In Fig.~\ref{figA2} we illustrate the graphs of $P(m)$ given by 
eq.~(\ref{pm2}) and corresponding numerical data.
There is a sharp peak at $m=1$ at all the densities and the 
distributions are exponential.
This means that large clusters of ants are rarely seen in this model.
Moreover, (\ref{pm2}) fits well with the numerical 
data for all $\rho > 0.5$, 
but it underestimates the simulations data at lower densities.

\begin{figure}[tb]
\begin{center}
\includegraphics[width=0.23\textwidth]{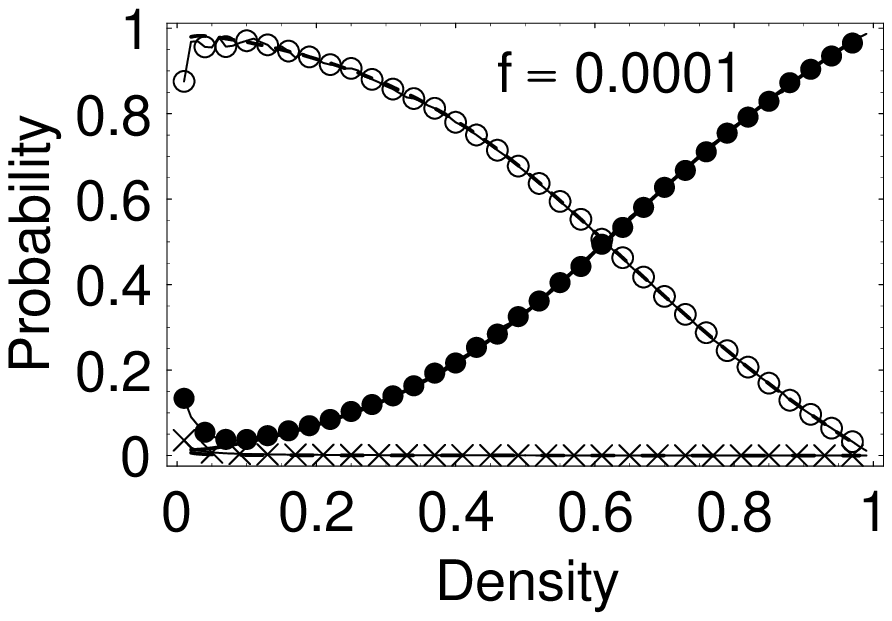}
\includegraphics[width=0.23\textwidth]{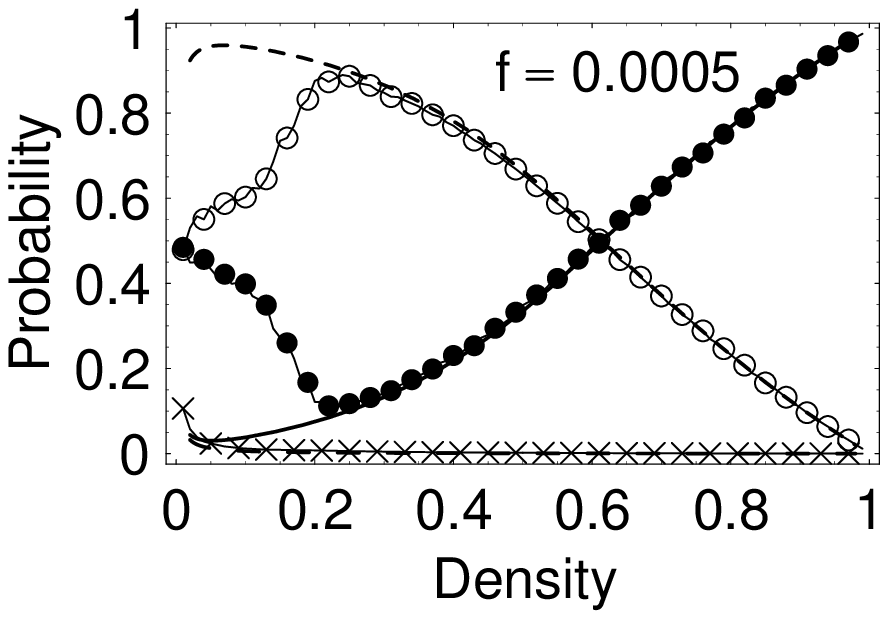}\\
\includegraphics[width=0.23\textwidth]{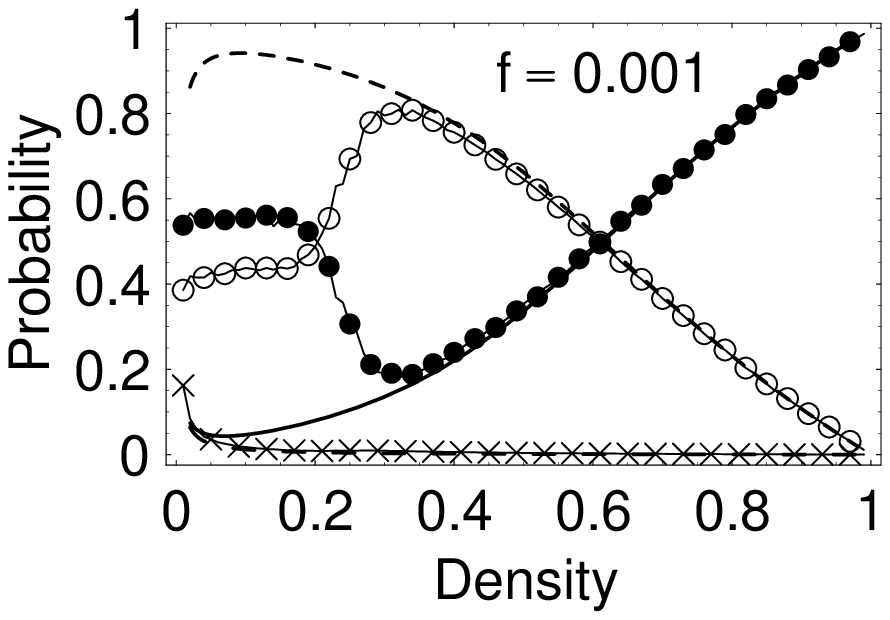}
\includegraphics[width=0.23\textwidth]{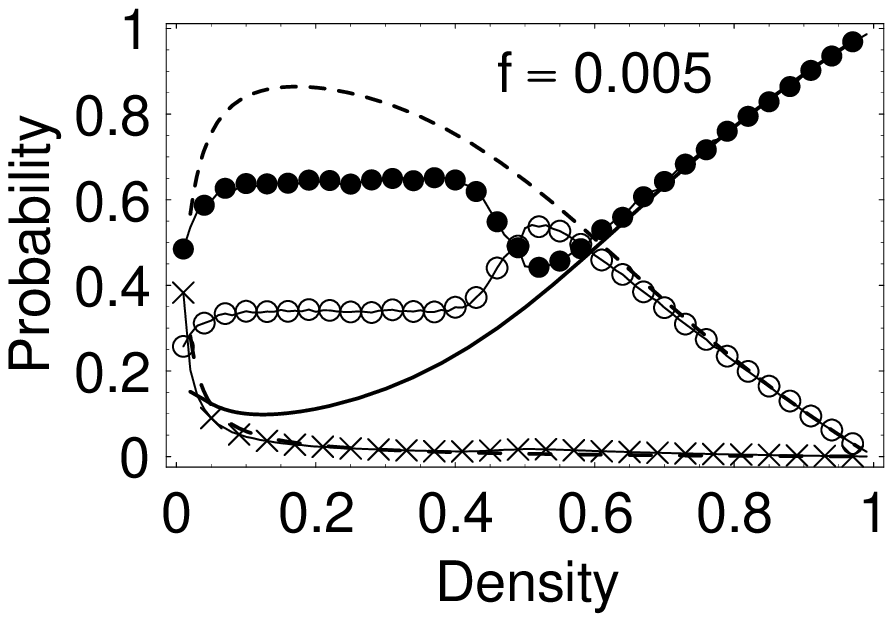}\\
\includegraphics[width=0.23\textwidth]{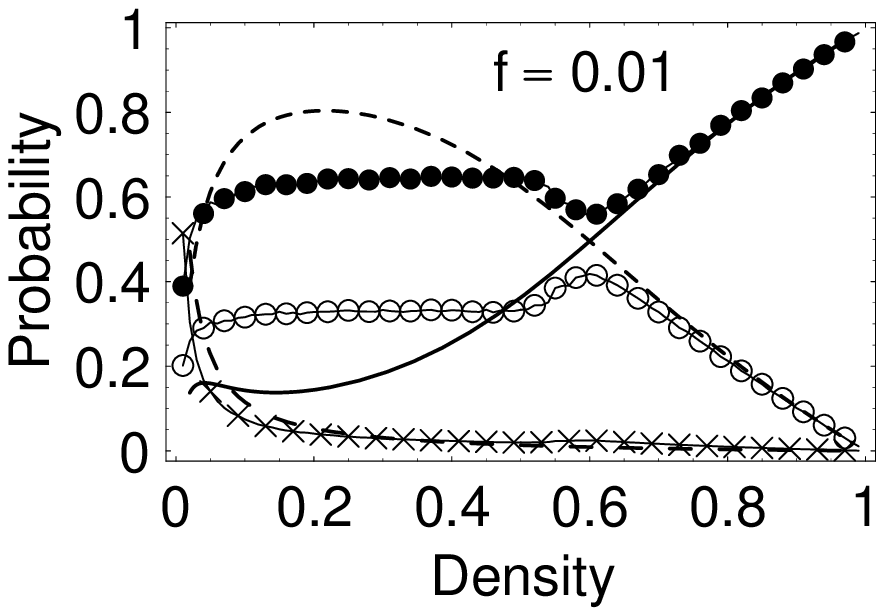}
\includegraphics[width=0.23\textwidth]{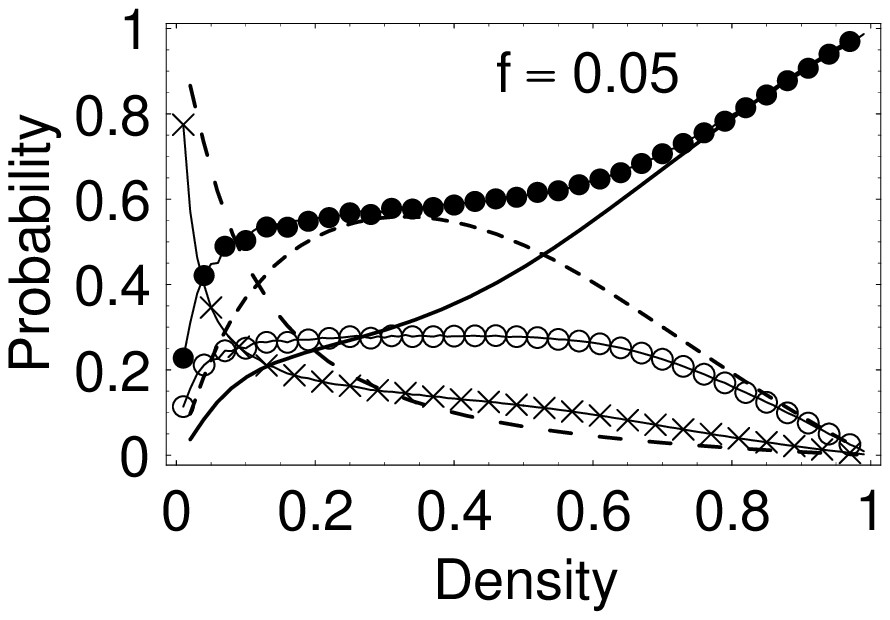}\\
\includegraphics[width=0.23\textwidth]{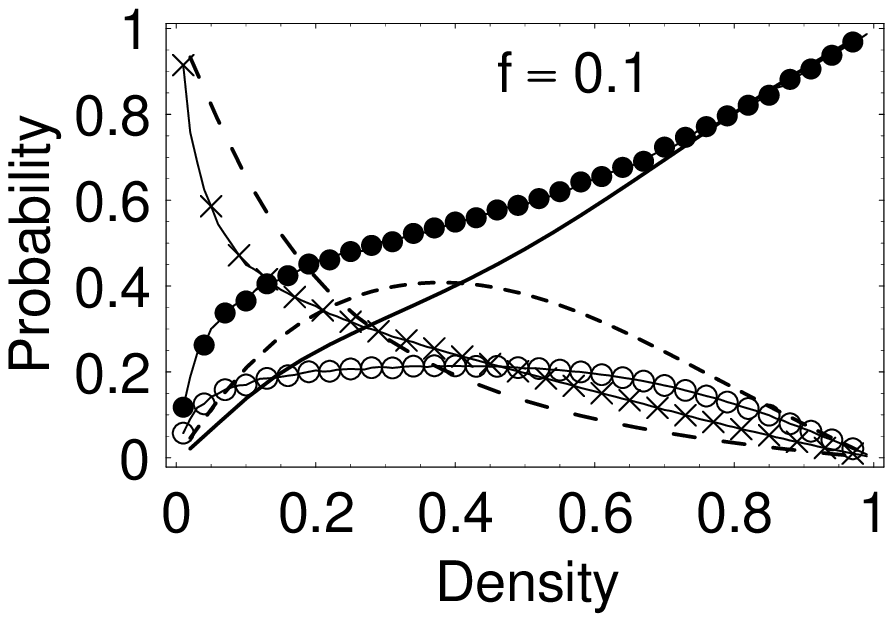}
\includegraphics[width=0.23\textwidth]{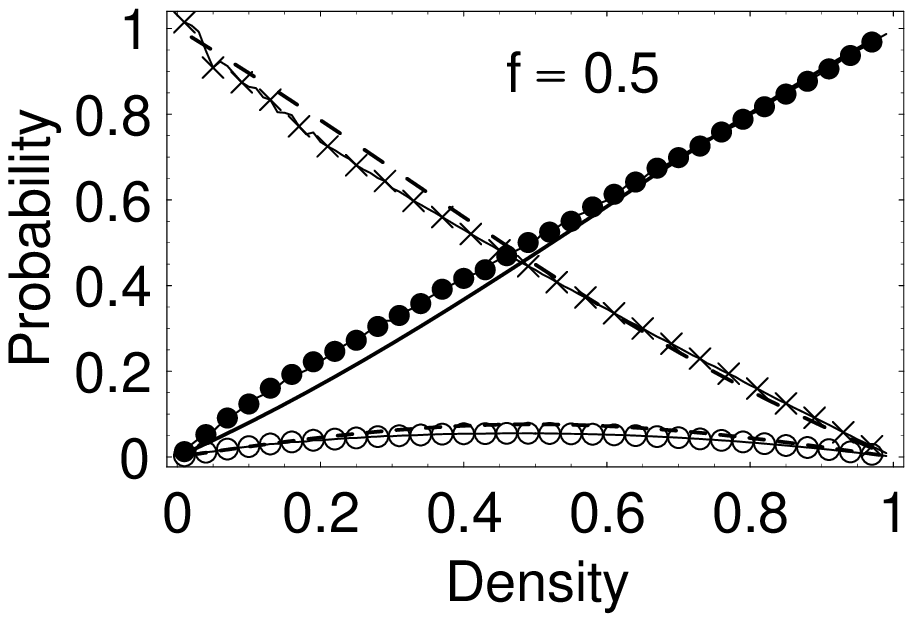}\\
\end{center}
\caption{Probability of finding an ant (solid curve), pheromone
(fine broken curve) and nothing (coarse broken curve) in front of an 
ant, with  parameters $f=0.0001,\cdots,0.5$, in the $(2+1)$-cluster 
approximation. 
Numerical data, obtained from computer simulations, are
also plotted (ants($\bullet$), pheromone but no ant($\circ$) 
and nothing($\times$)).
These figures show demonstrate how the predictions of the 
$(2+1)$-cluster approximation deviate from simulation data.}
\label{fig-A3}
\end{figure}

In order to get a better understanding of the microscopic structure 
of the stationary state, we also calculate the probabilities of finding 
an ant ($P_a$), pheromone ($P_p$) and nothing ($P_0$) in front of an ant:
\begin{eqnarray}
 P_a&=& 1-\frac{P(10)}{\rho},\label{ppprobs1}\\
 P_p&=& \frac{P(10)}{\rho(1-\rho)}
P\left(\begin{array}{c}0\\1\end{array}\right),\label{ppprobs2}\\
 P_0&=& \frac1\rho P(10)-
\frac{P(10)}{\rho(1-\rho)} 
P\left(\begin{array}{c}0\\1\end{array}\right).
\label{ppprobs3}
\end{eqnarray}
Note that the sum of these three probabilities is 1.
The results are shown in Fig.~\ref{fig-A3}.
We see that only for small and large values of $f$ (e.g., $f = 0.0001$ 
and $f > 0.1$) the results of the $(2+1)$-cluster approximation are in 
good quantitative agreement with the corresponding numerical results.

The results derived in this section indicate that the {\em microscopic}
cluster approximation is not able to capture the correlations which
lead to the sharp crossover observed for small evaporation
probabilities $f$. A systematic extension of this approximation scheme
is, in principle, possible and more correlations could be taken into
account. However, this approach will become quite cumbersome.
In Appendix~\ref{AppendB} we have also tried to extend the results
in this section by using the stochastic cluster approach \cite{mahnke},
but the results are not much improved.
In the following we, therefore, develope a {\em phenomenological}
mean-field theory which tries to capture the essential effects in
a simple way.

\section{\label{sec4}Homogeneous Mean-Field Theory (HMFT)} 
In this mean-field theory (MFT), let us assume
that all the ants move with the mean speed $V$ which depends
on the density $\rho$ of the ants as well as on $f$; although, to begin
with, the nature of these dependences are not known we'll obtain
these self-consistently.

Let us consider a pair of ants having a gap of $n$ sites in between.
We designate the leading ant of this pair as the lead ant (LA)
and the other as the following ant (FA). The probability that 
the site immediately in front of the FA contains pheromone is 
$(1-f)^{n/V }$. Here $n/V$ is just
the average time passed since the LA has dropped the pheromone.
Therefore, in this zeroeth level MFT, 
the effective hopping probability is given by
\begin{equation}
\eta_0 = Q (1-f)^{n/V} + q \left\{
1-(1-f)^{n/V}\right\}.
\label{lo1}
\end{equation}
In the mean-field approximation, we replace $n$ by the corresponding exact 
global mean separation $\langle n\rangle  = \frac{1}{\rho} - 1$ between 
successive ants, i.e.\ we are assuming the existence of a {\em homogeneous} 
state. Moreover, since $V_{{\rm max}} = 1$ the average speed $V$ 
is identical to the effective hopping probability, and we get the equation
\begin{equation}
 \left(\frac{\eta_0-q}{Q-q}\right)^{\eta_0}=(1-f)^{\frac1\rho -1}
\label{eqe0}
\end{equation}
which is to be solved self-consistently for getting $\eta_0$ as
a function of $\rho$ for a given $f$. 

Before solving the equation (\ref{eqe0}) numerically, note that this 
equation implies that, for {\it given} $f$, 
$\lim_{\rho \rightarrow 0} \eta_0 = q$;
this reflects the fact that, in the low-density regime, the pheromone
dropped by an ant gets enough time to completely evaporate before the
FA comes close enough to smell it. Equation (\ref{eqe0})
also implies that $\lim_{\rho \rightarrow 1} \eta_0 = Q$; this captures
the sufficiently high density situations where the ants are too close
to miss the smell of the pheromone dropped by the LA unless
the pheromone evaporation probability $f$ is very high. Similarly,
from the equation (\ref{eqe0}) we get, for {\it given} $\rho$,
$\lim_{f \rightarrow 1} \eta_0 = q$ and $\lim_{f \rightarrow 0}\eta_0 = Q$,
which are also consistent with intuitive expectations.

\begin{figure}[tb]
\begin{center}
\includegraphics[width=0.45\textwidth]{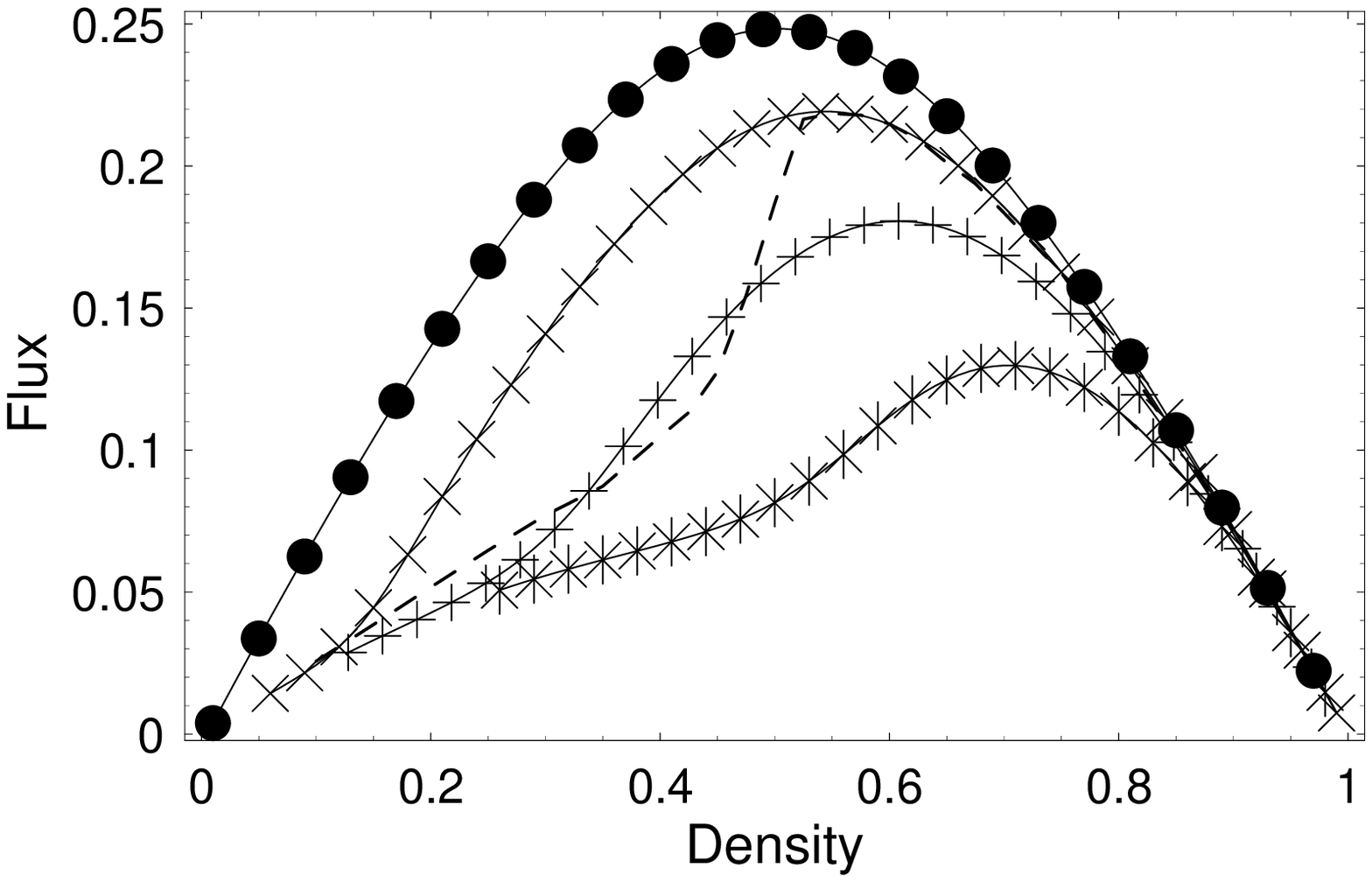}\\ 
\hspace{1cm}(a)\\
\vspace{0.5cm}
\includegraphics[width=0.45\textwidth]{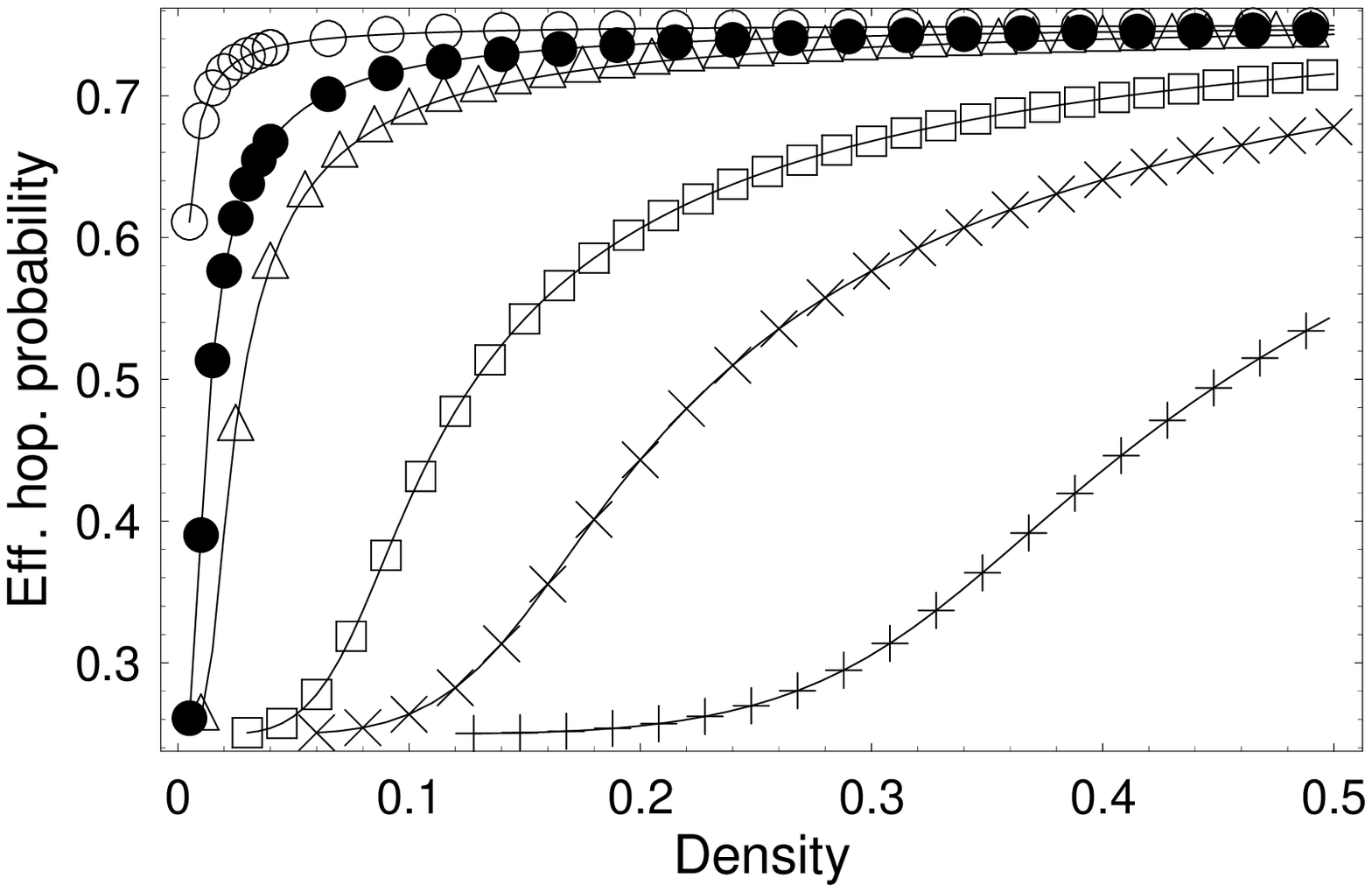}\\
\hspace{1cm}(b)
\end{center}
\caption{The fundamental diagram, obtained in the HMFT, is plotted against 
density in (a) while the corresponding effective hopping probability 
$\eta_0$ is shown in (b). 
The predictions of the HMFT are shown by the continuous curves; 
the same symbols in Figs.~\ref{fig-effh} and \ref{fig-hmft} correspond 
to the same values of $f$. The broken curve in (a), corresponding to 
the computer simulation data for $f = 0.005$ taken from 
Fig.~\ref{fig-effh}(a), highlights the limitations 
of the HMFT in making {\em quantitatively} accurate predictions.
}
\label{fig-hmft}
\end{figure}

The solutions of (\ref{eqe0}), calculated numerically by using Newton 
method, are plotted in Fig.~\ref{fig-hmft}(b) and the corresponding 
fundamental diagram is shown in Fig.~\ref{fig-hmft}(a). Clearly, the 
HMFT captures the {\it qualitative} features 
of the ant-trail model. However, there are significant {\it quantitative} 
differences between the predictions of this theory and the computer 
simulation data, especially, the sharp crossover around $\rho=0.5$ 
(Fig.~\ref{fig-effh}).  
One possible reason is that the HMFT assumes a rather homogeneous
stationary state. Therefore, in the following section we will 
develope an approximation scheme which emphasizes the formation
of a special kind of cluster in the steady state.

\section{\label{sec5}''Loose'' cluster approximation (LCA)} 
\begin{figure}[tb]
\begin{center}
\includegraphics[width=0.45\textwidth]{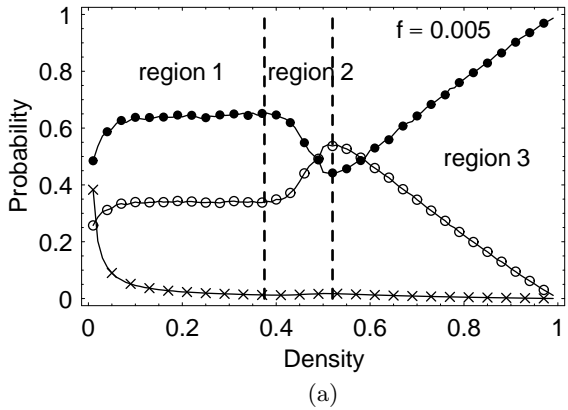}\\
\hspace{1cm}(a)\\
\vspace{0.5cm}
\includegraphics[width=0.45\textwidth]{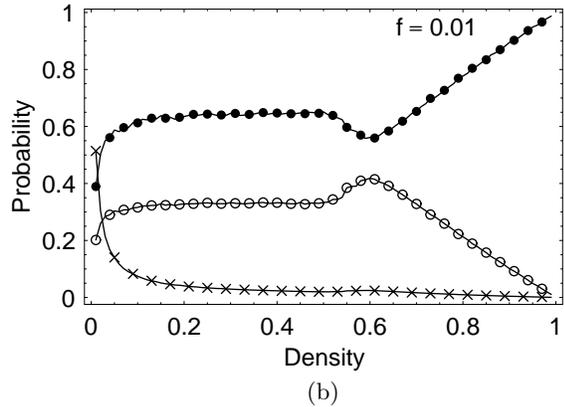}\\
\hspace{1cm}(b)
\end{center}
\caption{Numerical results for the probabilities of finding an 
ant($\bullet$), pheromone but no ant($\circ$) 
and nothing($\times$) in front of an ant are plotted against 
density of the ants.
The parameters 
are $f=0.005$ (in (a)) and $f=0.01$ (in (b)). See also Fig.~\ref{fig-A3}.
}
\label{fig-lca}
\end{figure}

Let us consider again the probabilities $P_a$, $P_p$, $P_{0}$ defined
in the previous section. For the purpose of clarifying some subtle 
concepts of "clustering" we replot these probabilities for only two 
specific values of $f$ in Fig.~\ref{fig-lca}; these 
data have been obtained from computer simulations of our ant-trail 
model. 

There is a flat part of the curves in Fig.~\ref{fig-lca} in the low 
density regime; from now onwards, we shall refer to this region as 
``region 1''. Note that in this region, in spite of low density of 
the ants, the probability of finding an ant in front of another is 
quite high. This implies the fact that ants tend to form a cluster.
On the other hand, cluster-size distribution (Fig.~\ref{figA2}), 
obtained from our computer simulations, shows that 
the probability of finding isolated ants are always higher than that 
of finding a cluster of ants occupying nearest-neighbor sites. 

\begin{figure}[tb]
\begin{center}
\includegraphics[width=0.5\textwidth]{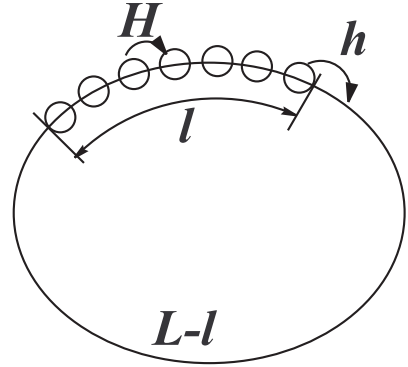} 
\end{center}
\caption{Schematic explanation of the loose cluster.
$H$ is the hopping probability of ants inside the loose cluster and
$h$ is that of the leading ant.
}
\label{fig-loose}
\end{figure}

These two apparently contradictory observations can be reconciled by 
assuming that the ants form ``loose'' clusters in the region 1.  The 
term ``loose'' means that there are small gaps in between successive
ants in the cluster, and the cluster looks like an usual compact
cluster if it is seen from a distance (Fig.~\ref{fig-loose}). In other 
words, a loose cluster is just a loose assembly of isolated ants.
Thus it corresponds to a space region with density larger than the
average density $\rho$, but smaller than the maximal density ($\rho=1$)
of a compact cluster.

Let us assume that the loose cluster becomes stationary after sufficient 
time has passed. Then the hopping probability of all the ants, except 
the leading one, is assumed to be $H$, while that of the leading one 
is $h$ (see Fig.~\ref{fig-loose}); 
the values of $H$ and $h$ are determined self-consistently, 
just as the effective hopping probability in the HMFT was estimated 
self-consistently in Sec.~\ref{sec4}. Before beginning the detailed 
analysis, let us consider the properties of $H$ and $h$.
If $f$ is small enough, then $H$ will be close to $Q$
because the gap between ants is quite small. On the other 
hand, if the density of ants 
is low enough, then $h$ will be very close to $q$ because 
the pheromone dropped by the leading ant would evaporate
when the following ant arrives there.

\begin{figure}[tb]
\begin{center}
\includegraphics[width=0.45\textwidth]{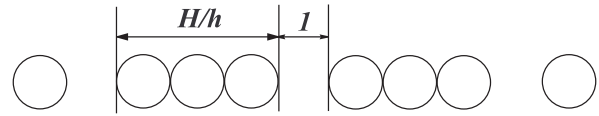} 
\end{center}
\caption{The stationary loose cluster.
The average gap between ants becomes $h/H$, 
which is irrelevant to the density of ants.}
\label{fig-avegap}
\end{figure}

Next we determine the typical size of the gap between successive 
ants in the cluster. 
We will estimate this by considering a simple time evolution beginning
with an usual compact cluster (with local density $\rho=1$) without any gap 
in between the ants. 
Then the leading ant will move forward by one site over the  
time interval $1/h$. This hopping occurs repeatedly and in the 
interval of the successive hopping, the number of the following ants
which will move one step is $H/h$. Thus, in the stationary state, 
strings (compact clusters) of length $H/h$, separated from each other 
by one vacant site, will produced repeatedly by the ants 
(see Fig.~\ref{fig-avegap}). 
Then the average gap between ants is 
\begin{equation}
 \frac{\left(\frac{H}{h}-1\right)\cdot 0+ 1\cdot 1}{H/h}=\frac{h}{H},
\end{equation}
which is independent of the density $\rho$ of ants. Interestingly, the 
density-independent average gap in the LCA is consistent with the 
flat part (i.e., region~1) observed in computer simulations 
(Fig.~\ref{fig-lca}).
In other words, the region~1 is dominated by loose clusters. 

Beyond region 1, the effect of pheromone of the last ant becomes
dominant.  Then the hopping probability of leading ants becomes large
and the gap becomes wider, which will increase the flow.  We call this
region as region 2, in which the ``looser'' cluster is formed in the
stationary state. It can be characterized by a negative gradient of
the density dependence of the probability to find an ant in front
of a cell occupied by an ant (see Fig.~\ref{fig-lca}).

Considering these facts, we finally
obtain the following equations for $h$ and $H$: 
\begin{eqnarray}
 \left(\frac{h-q}{Q-q}\right)^h=(1-f)^{L-l},\,\,
 \left(\frac{H-q}{Q-q}\right)^{H}=(1-f)^{\frac{h}{H}},
\label{simu}
\end{eqnarray}
where $l$ is the length of
the cluster given by
\begin{equation}
 l=\rho L +(\rho L -1)\frac{h}{H},
\end{equation}
and $\rho$ and $L$ are density and the system size, respectively.
These equations can be applied to the region 1 and 2, and
solved simultaneously by the Newton method.

Total flux in this system is then calculated as follows.
The effective density $\rho_{{\rm eff}}$ in the loose cluster is given by
\begin{equation}
 \rho_{{\rm eff}}=\frac{1}{1+h/H}.
\end{equation}
Therefore, considering the fact that
there are no ants in the part of the length $L-l$,
total flux $F$ is
\begin{equation}
 F=\frac{l}{L}f(H,\rho_{{\rm eff}}),
\end{equation}
where $f(H,\rho_{{\rm eff}})$ is given by
\begin{equation}
 f(H,\rho_{{\rm eff}})=\frac12\left(1-\sqrt{1-4H\rho_{{\rm eff}}
(1-\rho_{{\rm eff}})}\right).
\end{equation}

Above the density $1/2$, ants are assumed to be uniformly
distributed, in which a kind of MFA works well. 
We call this region as region~3. Thus we have three typical
regions in this model.
In region 3, the relation $H=h$ holds because all the gaps have
the same length, i.e.\ the state is homogeneous. Thus $h$ is determined by
\begin{eqnarray}
 \left(\frac{h-q}{Q-q}\right)^h&=&(1-f)^{\frac1\rho-1},
\label{old}
\end{eqnarray}
which is the same as our previous paper, and
flux is given by $f(h,\rho)$.
It is noted that if we put $\rho=1/2$ and $H=h$, then (\ref{simu})
coincides with (\ref{old}).

\begin{figure}[tb]
\begin{center}
\includegraphics[width=0.45\textwidth]{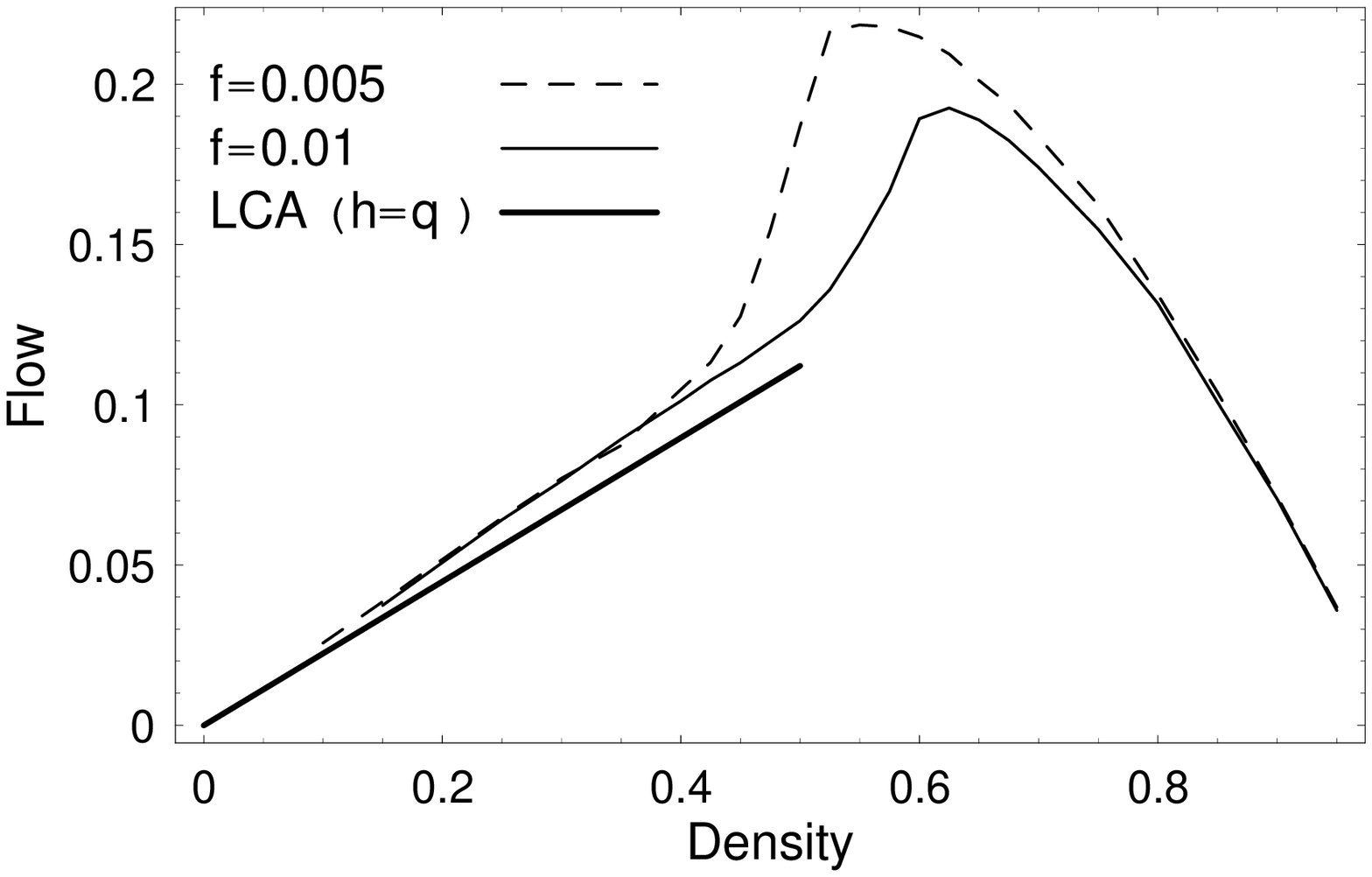}\\
\hspace{1cm}(a)\\
\vspace{0.5cm}
\includegraphics[width=0.45\textwidth]{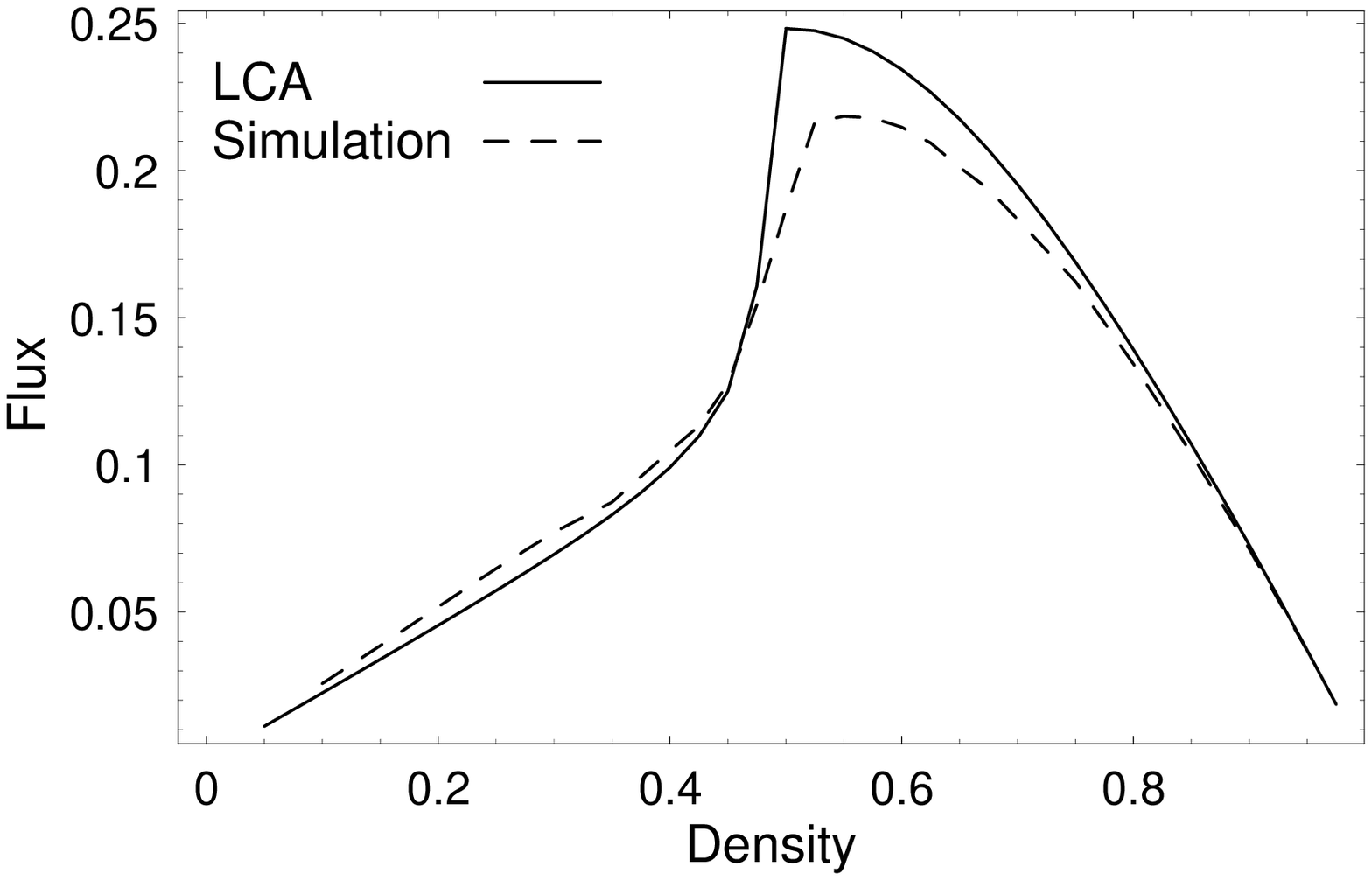}\\ 
\hspace{1cm}(b)
\end{center}
\caption{(a) Fundamental diagrams of the linear region (bold line)
together with 
numerical results with  parameters $f=0.005$ (broken curve) and 
$f=0.01$ (solid curve). 
(b) The fundamental diagram ($f=0.005$) of the combination 
of LCA and (\ref{old}) (solid curve). The broken curve is the
 numerical result for $f=0.005$. The system size is $L=350$.}
\label{fig-lin}
\end{figure}

We can focus on the region 1 by assuming $h=q$ in (\ref{simu}).
Under this assumption, we can easily see that 
the flux-density relation becomes linear.
In Fig.~\ref{fig-lin}(a), the two theoretical lines are almost the same, 
and the gradient of numerical results are also similar among
these values of $f$, which is quite similar to the theoretical one.
In Fig.~\ref{fig-lin}(b), the results obtained from (\ref{simu}) 
in the region $\rho\le 1/2$ are shown. 
Above this value of density, equation (\ref{old}) is used. 
The jointed curve fits quite well the numerical one.

\section{\label{sec6}Ant-trail model with random-sequential updating} 

In our earlier published work \cite{cgns} as well as in this paper 
so far we have considered only parallel updating of the states of 
the model system. However, in some models different updating schemes  
are known to give rise to nontrivial differences. For example, the 
correlations observed in the NaSch model \cite{ns} with parallel 
dynamics and $V_{{\rm max}}=1$ totally disappear when the parallel 
updating scheme is replaced by random sequential updating. In contrast, 
the updating 
scheme does not make much of difference in the bus route model 
\cite{loan,cd}. Therefore, in this section we examine the effects of 
replacing the parallel updating by random sequential updating, 
particularly on the unusual features of the fundamental diagram. 

In the ant-trail model with random sequential updating, the updating 
of the system is done the following way:
\begin{description}
\item{1)} A site is choosen randomly.
\item{2a)} If there is no ant, but a pheromone, at the chosen site this is 
allowed to evaporate with probability $f$.
\item{2b)} On the other hand, if there is an ant at the chosen site, the 
usual motion update is done (i.e., it cannot move forward if the site 
in front is occupied by another ant; otherwise, it moves forward with 
probability $Q$ or $q$ depending on whether the site in front contains 
or does not contain pheromone).
\item{3)} If the ant at the randomly chosen site moves forward, a pheromone 
is created at the new site without making any attempt to let the 
pheromone left behind in its old position (i.e., at the randomly chosen 
site) to evaporate.
\end{description} 

\begin{figure}[tb]
\begin{center}
\includegraphics[width=0.45\textwidth]{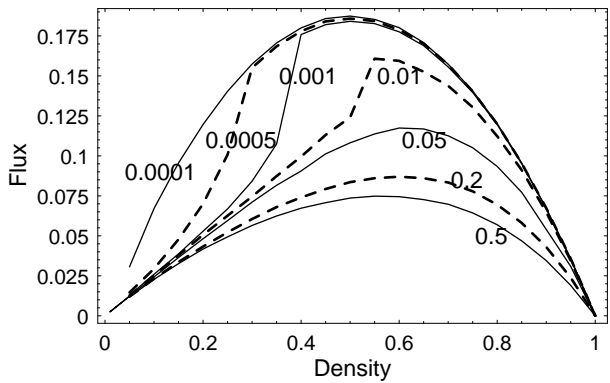}\\ 
\end{center}
\caption{The flux of the ants, in the ant-trail model with 
{\it random sequential updating}, plotted against their densities 
for the parameters $Q = 0.75, q = 0.25$. 
  }
\label{fig-rsu}
\end{figure}

The flux of the ants in this model is plotted against their 
density in Fig.~\ref{fig-rsu}; the qualitative features of the curves, 
including the sharp crossover from free to congested state, 
are similar to those in the original 
version of this model with parallel updating.
From these observations we conclude that, unlike the NaSch model, the 
correlations responsible for the non-monotonic variation of the average 
speed with the density of the ants are not artefacts of the parallel 
update scheme but genuine non-trivial features of the model. 

\section{\label{sec7}Concluding discussions}

A stochastic cellular automaton model of an ant trail, which we have 
proposed recently \cite{cgns}, has been investigated in detail, 
both analytically as well as numerically, in this paper. 
The model is characterized by two coupled dynamical variables,
representing the ants and the pheromone. The coupling leads to
surprising results, especially an anomalous fundamental diagram.
This anomalous shape of the fundamental diagram is a consequence of 
the non-monotonic variation of the average speed of the ants with 
their density in an intermediate range of the rate of pheromone 
evaporation. These unusual features of the ant-trail model have 
been analyzed in this paper using various analytical approaches 
and computer simulations. 

It is shown that the homogeneous mean-field approximations are 
able to capture some of the qualitative features observed in the 
computer simulations. However, these approximations cannot account 
for the quantitative data. Therefore, we have analyzed the 
spatio-temporal organization of the ants and pheromone in the 
stationary state. This provided some insights which we have utilized 
to develope a new scheme of calculations that we call ``loose-cluster 
approximation''. 

%

By studying appropriate correlation functions we were able to
distinguish three different regimes of density. At low densities (region 1)
the behaviour is dominated by the existence of loose clusters
which are formed through the interplay between the dynamics of
ants and pheromone. In region 2, occuring at intermediate densities,
the enhancement of the hopping probability due to pheromone is
dominant. 
Finally, in region 3, at large densities the mutual hindrance against 
the movements of the ants dominates the flow behaviour leading to a 
homogeneous state similar to that of the NS model.

We have seen that the observed effects persist for random sequential
updating. For this case we also expect that exact results can
be achieved by using the matrix-product technique \cite{derrida,derrida2}. 
Extensions of this model, including counterflow and random sequential
dynamics, will be reported in the future.

\begin{acknowledgments}
This work is supported, in part, 
by the Alexander von Humboldt Foundation, under the re-invitation 
programme for former Humboldt fellows (DC). 
\end{acknowledgments}
\appendix

\section{(2+1)-cluster approximation}
\label{AppendA}

In this appendix we provide details for the ``(2+1)``-cluster approximation 
scheme developed in Sec.~\ref{secClust}. There we have introduced the
eight dynamical variables (\ref{PSsigma}) which allow to take into
account correlations between occupation numbers of consecutive sites
and between occupation numbers of ants and pheromone.

These variables are not independent. Instead we can immediately write 
down the following six equations
\begin{eqnarray}
P(10)&=&P(01), \label{c1}\\
P\left(\begin{array}{c}1\\0\end{array}\right)&=&0, \label{c2}\\
 P(00)+P(01)+P(10)+P(11)&=&1, \label{c3}\\
P\left(\begin{array}{c}0\\0\end{array}\right)+
P\left(\begin{array}{c}0\\1\end{array}\right)+
P\left(\begin{array}{c}1\\0\end{array}\right)+
P\left(\begin{array}{c}1\\1\end{array}\right)&=&1, \label{c4}\\
P(00)+P(10)&=&1-\rho,\label{c5}\\
P\left(\begin{array}{c}0\\0\end{array}\right)+
P\left(\begin{array}{c}0\\1\end{array}\right)&=&1-\rho,\,\,\,\,\,\,\,\,\,\,
\,\,\,\label{c6}
\end{eqnarray}
where $\rho$ is the ant density. Equation (\ref{c1}) expresses the 
particle-hole symmetry condition while the equation (\ref{c2}) is 
a consequence of the definition of the model. The other equations
are known as Kolmogorov consistency conditions \cite{kolmo}.

We need more two equations in order to obtain the expression for all the
eight variables in (\ref{PSsigma}).
These are obtained by considering 
the master equations for, say, $P(00)$ and $
P\left(\begin{array}{c}0\\1\end{array}\right)$.
The master equation for $P(00)$ is given by
\begin{widetext}
\begin{eqnarray}
 \bar{P}(00)&=&\frac{P(00)}{P(00)+P(10)}P(00)
+\frac{P(10)}{P(00)+P(10)}(1-q_{{\rm eff}})P(00)
+\frac{P(00)}{P(00)+P(10)}P(01)\frac{P(10)}{P(10)+P(11)}q_{{\rm eff}}
\nonumber\\
&&
+\frac{P(10)}{P(00)+P(10)}(1-q_{{\rm eff}})P(01)
\frac{P(10)}{P(10)+P(11)}q_{{\rm eff}},
\label{mas1}
\end{eqnarray}
\end{widetext}
where $\bar{P}(00)$ is $P(00)$ of the next time step, i.e., at the 
time step $t+1$ while the probabilities on the right hand side refer 
to the time step $t$. The four terms in the r.h.s of (\ref{mas1})  
comprise all the configurations and processes that give rise to 
the configuration $(S_{j-1}S_j)=(00)$ in the next time step. Here we put 
$\bar{P}(00)=P(00)$ in (\ref{mas1}) in order to obtain the stationary 
solution for $P(00)$. Then we have
\begin{equation}
 P(00)=\frac{(1-q_{{\rm eff}})P(10)^2}{\rho-P(10)}.
\label{solmas1}
\end{equation}
Thus substituting (\ref{solmas1}) into (\ref{c5}) using (\ref{defF}), 
we obtain
\begin{equation}
 F^2-F+q_{{\rm eff}}\rho(1-\rho)=0.
\label{floweq}
\end{equation}

Similarly, the master equation for 
$P\left(\begin{array}{c}0\\1\end{array}\right)$
is given by
\begin{eqnarray}
 \bar{P}\left(\begin{array}{c}0\\1\end{array}\right)
&=&\frac{P(00)}{P(00)+P(10)} P\left(\begin{array}{c}0\\1\end{array}\right)
(1-f)\nonumber\\
&&+\frac{P(10)}{P(00)+P(10)} P\left(\begin{array}{c}0\\1\end{array}\right)
(1-q_{{\rm eff}})(1-f)\nonumber\\
&&+ P\left(\begin{array}{c}1\\1\end{array}\right)
\frac{q_{{\rm eff}} P(10)}{P(10)+P(11)}(1-f).
\label{mas2}
\end{eqnarray}
Using $P\left(\begin{array}{c}1\\1\end{array}\right)=\rho$, 
we obtain the stationary solution from (\ref{mas2}) as
\begin{equation}
 P\left(\begin{array}{c}0\\1\end{array}\right)
 =\frac{(1-f)F}{\displaystyle{f+\frac{1-f}{1-\rho}F}},
\label{solmas2}
\end{equation}
where we use the relation 
\begin{equation}
 F=q_{{\rm eff}} P(10) = P(10)\frac{qP\left(\begin{array}{c}0\\0
\end{array}\right)+
QP\left(\begin{array}{c}0\\1\end{array}\right)}{
P\left(\begin{array}{c}0\\0\end{array}\right)+
P\left(\begin{array}{c}0\\1\end{array}\right)}.
\label{defF2}
\end{equation}

From (\ref{defF2}) and (\ref{solmas2}) we have
\begin{equation}
 q_{{\rm eff}}=q+(Q-q)\frac{(1-f)F}{(1-\rho)f+(1-f)F}.
\label{eta}
\end{equation}
Finally, substituting (\ref{eta}) into (\ref{floweq})
we obtain the cubic equation (\ref{flowfin}) for the flux $F$. 

We can also calculate the distribution of cluster sizes
defined in Sec.~\ref{secClust}.
We can write it as 
\begin{eqnarray}
 P(m)&=&\frac1C P(01)\left(\frac{P(11)}{P(10)+P(11)}
\right)^{m-1}\frac{P(10)}{P(10)+P(11)}\nonumber\\
&=&\frac1C \frac{P(10)^2}{\rho}\left(1-\frac{P(10)}{\rho}
\right)^{m-1}.
\end{eqnarray}
Here $C$ is determined through the normalization condition 
$\sum_{m=1}^{L}P(m)=1$,
where $L$ is the system size, as
\begin{equation}
 C=P(10)\left\{  1-\left(1-\frac{P(10)}{\rho}\right)^{\rho L}\right\}.
\end{equation}
Thus $P(m)$ is given by (\ref{pm2}).

Let us also consider the 
the probability of finding an ant, pheromone and nothing
in the front site discussed in Section~\ref{sec5}.
The probability of finding an ant is simply given by $P(11)$. 
The probability of finding pheromone without an ant, and
that of nothing are given respectively by
\begin{equation}
P(10)\frac{P\left(\begin{array}{c}0\\1\end{array}\right)}{
P\left(\begin{array}{c}0\\0\end{array}\right)+
P\left(\begin{array}{c}0\\1\end{array}\right)},\,\,\,\,
P(10)\frac{P\left(\begin{array}{c}0\\0\end{array}\right)}{
P\left(\begin{array}{c}0\\0\end{array}\right)+
P\left(\begin{array}{c}0\\1\end{array}\right)}.
\end{equation}
Normalizing these quantities by dividing $\rho$, we obtain each 
probability by only using $P(10)$ and 
$P\left(\begin{array}{c}0\\1\end{array}\right)$ as given in
equations (\ref{ppprobs1})--(\ref{ppprobs3}).

\section{Stochastic cluster approximation}
\label{AppendB}
Let us extend the analysis in the Appendix \ref{AppendA} following
the approach used in analyzing the stochastic car cluster model proposed 
in \cite{mahnke}.
In the model, one cluster of cars is assumed to exist in the background 
of stationary uniform flow, while in Appendix \ref{AppendA}
we only consider the uniform flow to derive flux of ants, and neglect 
the clustering effect.
The cluster-size distribution $P(m)$ was derived as
the stationary solution of its master equation in \cite{mahnke}. 
However, since we have already obtained $P(m)$, we will use (\ref{pm2})
instead of considering the master equation.
The flux in cluster is considered to be zero, thus
total flux in a given configuration 
of this system is given by $0\times (m-1)/L + F_m\times (1-(m-1)/L)$ if
$m$-size cluster exists.
Here $F_m$ represents 
the uniform flux under the existance of $m$-size cluster,
which is defined by using (\ref{flowfin}) as
\begin{equation}
 F_m^2-F_m+\rho_m(1-\rho_m)\left\{
q+\frac{(Q-q)(1-f)F_m}{(1-\rho_m)f+(1-f)F_m}
\right\}=0,
\end{equation}
and $\rho_m$ is given by
\begin{equation}
 \rho_m=\frac{\rho-(m-1)/L}{1-(m-1)/L}.
\end{equation}
In these equations we take into account that
the density of the uniform flow is reduced due to the existance of
$m$-size cluster.

Thus the flux of this stochastic cluster approximation is finally given by
\begin{equation}
 F(\rho)=\sum_{m=1}^LP(m)\left(1-\frac{m-1}{L}\right)F_m.
\label{finF}
\end{equation}

Note that if only the first term on the r.h.s of (\ref{finF}) is 
retained and all the other terms are dropped the expression for 
$F(\rho)$ reduces to the the fundamental diagrams obtained in the 
$(2+1)$-cluster approximation (and plotted in Fig.~\ref{fig-A1}). 
We have evaluated (\ref{finF}) numerically by using the distribution 
(\ref{pm2}), but the results are almost the 
same as Fig.~\ref{fig-A1} 
and, therefore, not shown here. This, however, is not surprising 
in view of the fact that the r.h.s of (\ref{finF}) is dominated by 
$m = 1$ because of the sharp peak of $P(m)$ at $m = 1$.


\end{document}